\documentclass[10pt,journal,compsoc]{IEEEtran}
\ifCLASSOPTIONcompsoc
  \usepackage[nocompress]{cite}
\else
  \usepackage{cite}
\fi

\usepackage{pgf-pie}
\usepackage{pgfplots}
\usepackage{float}
\usepackage{tikz}
\usepackage{cite}
\usepackage{url}
\usepackage{tabularx}
\usepackage{array}
\usepackage{multirow}
\ifCLASSINFOpdf
\else
\fi
\pgfplotsset{compat=1.14}
\begin{document}
\title{Indian EmoSpeech Command Dataset: A dataset for emotion based speech recognition in the wild. }

\author{Subham\,Banga,
       Ujjwal\,Upadhyay,
       Piyush\,Agarwal,
       Aniket\,Sharma
       and~Prerana\,Mukherjee% <-this % stops a space
\thanks{

S. Banga and P. Agarwal are with the Department of Information Technology, Bharati Vidyapeeth's College of Engineering, Paschim Vihar, New Delhi, 110063 India. (e-mail: subhambanga26@gmail.com and me@ipiyush.com)

U. Upadhyay and A. Sharma are with the Department of Computer Science, Bharati Vidyapeeth's College of Engineering, Paschim Vihar, New Delhi, 110063 India. (e-mail: ujjwalupadhyay8@gmail.com and aniket965.as@gmail.com)

P. Mukherjee is with the Department of Computer Science, Indian Institute of Information Technology, Sri City, Andhra Pradesh, 517646 India. (e-mail: prerana.m@iiits.in) 

 S. Banga, P. Agarwal, U. Upadhyay and A. Sharma contributed equally \& their names are in random order.

}}

\IEEEtitleabstractindextext{%
\begin{abstract}
Speech emotion analysis is an important task which further enables several application use cases. The non-verbal sounds within speech utterances also play a pivotal role in emotion analysis in speech. Due to the widespread use of smartphones, it becomes viable to analyze speech commands captured using microphones for emotion understanding by utilizing on-device machine learning models. The non-verbal information includes the environment background sounds describing the type of surroundings, current situation and activities being performed. In this work, we consider both verbal (speech commands) and non-verbal sounds (background noises) within an utterance for emotion analysis in real-life
scenarios. We create an indigenous dataset for this task namely "Indian EmoSpeech Command Dataset". 
 It contains keywords with diverse emotions and background sounds, presented to explore new challenges in audio analysis. We exhaustively compare with various baseline models for  emotion analysis on speech commands on several performance metrics. We demonstrate that we achieve a significant average gain of 3.3\% in top-one score over a subset of speech command dataset for keyword spotting. 
\end{abstract}

% Note that keywords are not normally used for peer review papers.
}


\maketitle

\IEEEdisplaynontitleabstractindextext

% \fi
\IEEEpeerreviewmaketitle

\IEEEraisesectionheading{\section{Introduction}\label{sec:introduction}}

\IEEEPARstart{A}{udio} and voice is a critical part of communication. It contains non-verbal information that accentuates, emphasizes and in some cases negates the content of the message. The tone of speech, background voices, and sounds divulge information about the primary speaker's current emotional status and their environment. Traditionally, the audio analysis had been focused primarily on speech recognition \cite{7472618,albanie2018emotion,kadiri2015analysis, mirsamadi2017automatic,trigeorgis2016adieu} and identification of the speaker and/or their demographic information \cite{DBLP:journals/corr/RichardsonRD15}. In recent years, however, several research studies have been formulated around the recognition of unsafe and abnormal events \cite{6327995}.

Event recognition via audio medium has attracted a lot of research effort that is aimed at developing machines to mimic human-like capability to identify and make correlations between certain sounds and events. The recognition models heavily depend upon processing audio signals from the environment and emotion of the primary speaker\cite{xu2017two,xie2019speech,deng2017semisupervised}. These models can benefit from an improved speech analysis that can be attained through a diverse dataset which includes various emotion and intensity levels in speech. In this paper, we present the Indian Emospeech command dataset, an audio dataset with diverse emotions and keywords infused with different background noise audio events. Emotion can be an important parameter for audio analysis, according to a research conducted by Polzin and Waibel\cite{polzin1998detecting}, the accuracy drops slightly for speech with different emotions if the training dataset only contains neutral speech samples or do not have enough diversity of emotions. The development of such audio event recognition model demands a large amount of labeled data\cite{lotfian2019curriculum}. Even semi-labeled data proves inadequate. Recently, the creation of datasets for research purposes is being done by trying to exploit the public audio archives available on the web\cite{45857}. High quality labeled data is not currently available for audio analysis. We envision that this dataset can serve as a pioneering work in potential distress analysis using audio signal analysis with on-device machine learning models.

To the best of authors knowledge, the Indian EmoSpeech Command Dataset is the first attempt to create a diverse dataset for audio event recognition. Indian EmoSpeech Command Dataset aims to provide on average 800-1000 audio samples to illustrate each emotion class. This dataset is intended to cater to the current requirements in security-related applications. Audio samples for each emotion and keyword are quality-controlled and manually-annotated as described in Sec. \ref{sec: Quality Control}.
Thus, Indian EmoSpeech Command Dataset offers thousands of organized and labeled high-quality speech samples with diverse emotions. In this paper, we report the current version of the Indian EmoSpeech command dataset, consisting of 6 "keywords": \textit{help, bachao\footnote{\textit{Bachao} means to safeguard in Hindi dialect}, stop, go, yes, no} with 4 "emotions": \textit{calm, happy, angry, fearful}. 

In view of above discussions, the key contributions in the paper can be summarized as,
\begin{itemize}
    \item We curate an indigeneous dataset for emotion detection in speech commands namely "Indian EmoSpeech command dataset". It has diverse set of emotions in audio keywords which enables emotion prediction during
    speech analysis. We make the dataset robust against background noises to sustain the audio analysis in crowded and noisy environments. 
    % We capture the Emotions in audio keywords: Presence of different emotions in our audio samples will help improve the speech analysis. 
    \item We perform a comprehensive analysis on keyword detection in varying conditions such as with the presence of background noise, varying emotions and varying environmental conditions. We compare the results with benchmark Speech Command dataset and show a significant performance gain of 3.3\% in top-one score for keyword spotting. 
    % Robustness to background audio: Different background noises in our audio samples allow it to work well in crowded and noisy areas.
    % Potential distress signals using additional audio features

\end{itemize}

The paper is organized as follows. In Sec. \ref{sec:motivate}, we provide the motivation behind this attempt of Indian EmoSpeech dataset curation. In Sec. \ref{sec:prop}, we define the characteristic properties associated with the dataset. In Sec. \ref{sec:collect}, we describe the details related to the dataset collection. In Sec. \ref{sec:related}, we provide relevant details on other contemporary datasets in this domain. In Sec. \ref{sec:eval}, we outline the evaluation results and analysis. In Sec. \ref{sec:appli}, we provide the potential applications or use cases of this dataset followed by the future work in this regard. Finally, we provide the conclusion in Sec. \ref{sec:conclu}.


\section{Motivation}
\label{sec:motivate}
 Nowadays voice interfaces rely on audio commands and keyword spotting to initiate an interaction. For instance, you might say "Ok Google" or "Hey Siri"\cite{heysiri} to interact with the voice assistant running on your smartphones. It is impractical to run a cloud-based service for the initial keyword detection since it would require sending audio data over the web all the time. This would not only increase the latency, but implementation and maintenance costs would also escalate. Such an approach also introduce several privacy risks for the user as each speech command is sent to the server. Therefore, Keyword spotting, Emotion detection is done locally on the mobile device with the help of edge computing models which is more reliable and cost effective.\newline
 Speech is the primary means of communication amongst humans, and if confined in meaning to the explicit verbal content of what is said, it does not, by itself, carry all the information that is required to be conveyed. Additional information includes vocalized emotion, if edge models could identify emotions, background events apart from just keywords then they could be used to create a system that not only invokes a smart assistant but could trigger an alert by analyzing audio input features and may also provide personalized triggering option to user for different emotions.\newline
 Hence, the audio interface can be utilized for multiple tasks than just to instantiate a query or command for your phone. Other than starting an interaction with the voice-based assistant's keyword spotting, audio commands play an important role in other areas such as safety and security of the user. Since everyone carries around a smartphone equipped with microphones, using audio input to ensure the safety and security of the user is a pragmatic approach. 
Following differences justify that the task of emotion infused keyword spotting with background events is quite different from generic keyword spotting in emotion and speech recognition models:
\begin{itemize}
    \item These models must be low memory footprint (compact) and involve less computation for emotion recognition, keyword spotting, and background analysis.
    
    \item False positives should be minimized for both emotion and keyword recognition.
    \item Most of the speech input will be unrelated which should not trigger the event.
    \item Each emotion must be distinguishable from other emotions for the same keyword or small speech phrase.
    \item Emotion and keyword recognition should be sustainable in noisy background surroundings.
\end{itemize}
Indian Emospeech command dataset aims to target the abovementioned requirements of creating edge computing models for emotion infused keyword detection.

\section{Properties of Indian EmoSpeech Command Dataset}
\label{sec:prop}

\subsection{Scale} 
\begin{figure}[ht]
    \centering
    \begin{tikzpicture}
\begin{axis}[
    ybar stacked,
    bar width=15pt,
    nodes near coords,
    enlargelimits=0.15,
    legend style={at={(0.5,-0.20)},
      anchor=north,legend columns=-1},
    ylabel={No of Samples},
    symbolic x coords={Help, Bachao, Yes, No, 
        Stop, Go},
    xtick=data,
    ymin=0,
    % x tick label style={rotate=45,anchor=east},
    ]
\addplot+[ybar] plot coordinates {(Help,341) (Bachao,318) 
  (Yes,324) (No,262) (Stop,272) (Go,363)};
\addplot+[ybar] plot coordinates {(Help,445) (Bachao,437) 
  (Yes,156) (No,153) (Stop,154) (Go,193)};
\addplot+[ybar] plot coordinates {(Help,188) (Bachao,74)
  (Yes,198) (No,204) (Stop,201) (Go,242)};
\addplot+[ybar] plot coordinates {(Help,96) (Bachao,92) 
  (Yes,199) (No,168) (Stop,161) (Go,167)};
\legend{\strut Calm, \strut Fearful, \strut Angry, \strut Happy}
\end{axis}
\end{tikzpicture}
    \caption{Emotion-Wise distribution of keywords}
    \label{fig:emo_dist}
\end{figure}
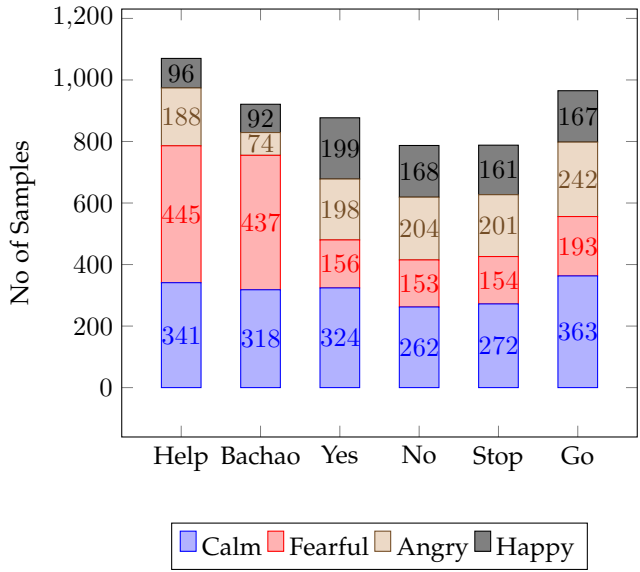

Indian EmoSpeech Command Dataset aims to provide the most comprehensive and diverse set of audio samples captured with different emotions, environments and for different keywords. The current collection consists of 8K audio samples. The dataset distribution based on different keywords is represented in Fig. \ref{fig:dist_keyword}. The distribution of keywords based on emotion is shown in Fig. \ref{fig:emo_dist}.
\begin{figure}[ht]
\centering
\begin{tikzpicture}
\pie{14.4/Background, 13.8/Help, 12.2/Bachao, 11.3/Yes, 10.2/No, 10.2/Stop, 12.5/Go, 15.4/Unknown}
\end{tikzpicture}
\caption{Distribution of keywords}
\label{fig:dist_keyword}
\end{figure}
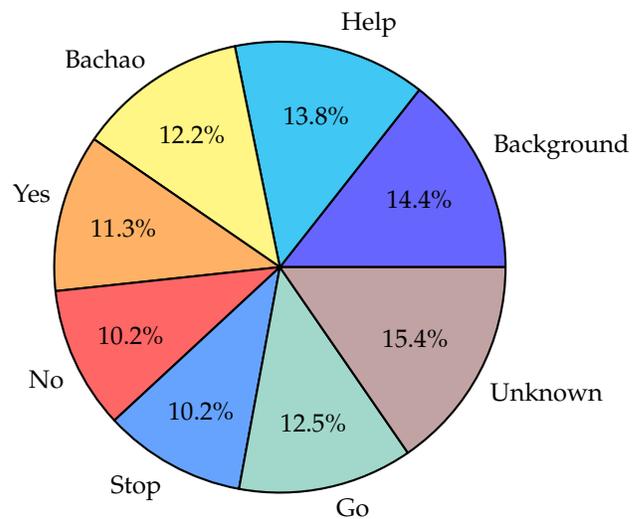

\subsection{Emotions}
% Describing in details about emotions
The audio samples are organized based on 4 different emotions: calm, happy, angry, fearful. These 4 different emotions can help identify the emotional state of a person. The information regarding emotion can be associated with speech and therefore help in measuring the distress a person might be in. Such a state can be useful for different applications. One of the main assets of the Indian Emospeech Command Dataset is that for each emotion, the data has been collected on different backgrounds and accents. Emotions vary from person to person so with each person we have captured a variation of an emotion that is unique in itself and hence help the model trained on it to generalize better.

\subsection{Diversity}
The Indian EmoSpeech Command dataset is diverse in many aspects. Some of them are listed below:
\begin{itemize}
    \item The emotions that are captured make it diverse in each keyword.
    \item Number of speakers that participated in this crowd-sourcing dataset is 250. 
    % in this exercise.%correct the number
    \item The real-world background noise such as the honking of a car, the noise of kids playing in a park, sounds recorded in temples, cafe, etc, makes it more diverse than any other audio dataset.
    \item Different Indian accents were captured in the audio.
    \item Collection point scattered across different landscapes in India.
\end{itemize}

\section{Collection}
\label{sec:collect}
\subsection{Requirements} \label{sec:requirements}
Indian Emospeech Command Dataset consists of audio samples that are recorded for 3 seconds. In order to collect this data, we record the samples in real-life environments. Recording the audio samples within studios would have defeated the purpose of having samples with real-life background sounds. The recorded data can be used for several scenarios where environmental noises play a crucial role. To record audio, an application was created that works on mobile devices, for contributors to contribute their voice with real-life environments.

The dataset aims to contain samples of diverse accents/dialects with the focus on an Indian accent. The rolling out of application was monitored through an anonymous analytic tool to allow the curators to get as much as diversity in people as possible. The aim was to allow a person to give as much as possible samples to the system.

\noindent The core requirements on the quality of the collected data were as follows:

\begin{itemize}
\item Each sample needs to be precisely 3000ms (milliseconds) long.
\item All audio files should be compressed using a lossless coding scheme.
\item There should be no fall in the audio level of recording by the microphone.
\item Single channel was used for collection of the data. Mobile devices generally have mono microphone only.
\item The rate of a sample of the recording should be 16000Hz. 
\item There should not be any echo in the recorded audio.

\end{itemize}

With the above requirements, the most important requirement of data collection was peer-reviewing the collected audio samples. The dataset was manually reviewed to maintain the quality of the data.

\subsection{Keywords Choice}
This dataset consists of 6 keywords: help, bachao (Hindi translation of help), stop, go, yes, no. These keywords were chosen for a variety of reasons. Firstly these keywords can be used in many different tasks. For instance, keywords help and bachao, they are useful for sending SOS signals and triggering panic events. While all other keywords that are yes, no, stop, go are also loosely coupled with these tasks. But these keywords can also be used for some other tasks involving speech command recognition. For example, stop and go can be used in instructing the autonomous system to function as per user requirement. The same analogy can be provided in case of keywords yes and no. Thus, the selected keywords present a diverse set of use cases which can be solved using other metadata information that is embedded in this dataset such as emotion and background sound along with audio based location description.

\subsection{Implementation}
During implementation, the primary focus is to meet all the requirement specifications so that we can make it easy for reviewers to review these samples. The system should be able to enforce the duration of recorded samples and no manual or semi-automated process should be needed to meet those requirements after recording samples. To collect this dataset with all these requirements, a web application was created to record audio samples\cite{quickspeech}, PCM (Pulse code modulation) encoded, with the help of open-source software and Media Recorder\cite{mrAPI} API.
The implementation began with first developing an android application for in-house generation of audio samples, reviewing the quality and building up the standards. After this, we worked on a web application that can make it easy for everyone to contribute to the data collection process as well as allow people to ensure their right to privacy is not inflicted upon.
To ensure privacy consent, we ask the user to agree to our privacy policy. On agreeing, a token is created for the whole session of the app. To release data in the public domain, any detail that can be linked with the identity of a person is removed from the label and metadata information. The standards for recording the audio were the same in both the applications. The web application was hosted on Firebase\cite{firebase}, a cloud based backend service by Google, which also provided the SSL certification thus making the application more secure.
To make the application more intuitive, we made a game-like experience. The game allows a person a window of 3 minutes on starting the game and each contribution of audio samples is rewarded with one point. The person can aim for a higher score. In each game, a user sees a word and an emotion on the screen. On pressing the record button, the user has to speak the word with the required emotion that is displayed there. 

\subsection{Quality Control} \label{sec: Quality Control}

This dataset aims to provide good quality audio samples that meet all requirements specified in Section \ref{sec:requirements}. The last phase on Quality control was peer-reviewing, if human reviewers can't tell the word in the audio, it is rejected straight away. Each sample is reviewed by two persons first, and all doubt cases were further sent to reviewing by more reviewers. In the same case with emotions in the audio, human reviewers are also responsible for matching the emotion in audio with a set of quality samples. These quality samples were produced in the initial phase by the whole team.
There were two interfaces for the dataset collection: i) the android application and ii) web-based application. The aim of the native android application was to start creating a quality dataset manually that will set the base of a future dataset that will be collected by the web-based application.

The service to collect the audio sample had a strict timer of 3000ms with 200ms of window to ensure that there is ample time for the machine to encode and save the recorded audio. With this process, the range of file size came 90 KBs - 120 KBs. The pipeline from web application monitored the samples that are coming from a crowd sourcing web application. Files that were too large or too small are straight away moved from storage to archive. This range takes consideration of all these pointers:
\begin{itemize}
    \item Lossless compression
    \item Noise suppression
    \item Echo cancellation
\end{itemize}

\subsection{Manual Review}
The audio data-set has a high chance of being a victim of technical issues when data collection is done through crowd-sourcing. These technical issues may include malfunctioning of microphones, corruption of the audio file or it may be due to some encoding issues. In order to counter such problems, we relied on manual reviewing of the data-set. Each audio was checked by at least 1 person. The process of verifying the label was done in the following manner:
\begin{itemize}
  \item Firstly the clip was played for the worker.
  \item Then the labels were shown to the worker.
  \item If the worker found the label to be incorrect then its label is changed by the worker. If the audio was found to be corrupt, then it is deleted. 
  \item The incorrectly labeled clip is then added to the review list so that it can be checked again.
\end{itemize}
This process helped us curate a data-set that is of high quality and has very few or no errors.

\subsection{Release}\label{Release}
The Indian EmoSpeech Command dataset is released in a way that gives the user ways to solve at least 4 problems. Each clip is located in a folder containing the place it was recorded from such as railway station, park, college, house, etc. 
\\Then each clip is named using the following convention:- 
\\\\\textbf{keyword}-\textbf{background}-\textbf{emotion}-timestamp.wav
\\\\For augmented data a random number is also appended after the timestamp. So its naming is as follows:-
\\\\\textbf{keyword}-\textbf{background}-\textbf{emotion}-timestamp-random.wav
\\\\This dataset is released for non-commercial purpose on: \begin{center}\fbox{\url{https://emo-speech.web.app/}}\end{center}
.\\Any Person having email id provided by an educational institution can send a request with the purpose of dataset usage, and the request will be verified within 2 days. If access is granted to the user, he/she can download the dataset within 7 days of acceptance email.

\section{Related Datasets}
\label{sec:related}

\begin{table*}[!ht]
\begin{tabular}{ ||p{3.5cm}|p{2cm}|p{2cm}|p{2cm}
|p{2cm}|p{2cm}|p{2cm}||}
 \hline
 \multicolumn{7}{||c||}{Datasets} \\
 \hline
 \hline
 Parameters &Indian EmoSpeech Command&Speech Command& Mozilla Common Voice& LibriSpeech& Ravdees& Mivia\\
 \hline
 Defined Keywords   & Yes    &Yes&   No&
 No&  No  & No\\
 Specific Sentences &No & No&  Yes& Yes& No& No\\
 Existence of Emotion   &Yes & No&  No& No& Yes& No\\
 Natural Background Noise&  Yes  & No& No& No& No& Yes\\
 Location Tagged Audio& Yes  & No & No& No & No &No\\
 No. of Samples& 8K  & 30K&39K & 1000hrs& 7K & 6K\\
 \hline
\end{tabular}\\
\caption{Comparison amongst different available speech datasets}
\label{table:datasets}
\end{table*}

\textbf{Speech Command dataset}\cite{scdv2} is a limited vocabulary speech recognition dataset. It is a collection of 30 keywords and a class for background noise. It was created by the TensorFlow and AIY(Artificial Intelligence Yourself)teams. The dataset has 65,000 one-second long utterances of 30 short words, by thousands of different people, contributed by public members through the AIY website. It is released under a Creative Commons by 4.0 license.\\
\textbf{Mozilla Common Voice}\cite{mcv} is a crowdsource dataset. It currently contains about 39000 voices. It contains over 780 hours worth of voice samples. Their voice sample collection platform asks the volunteers to read a specific sentence that gets validated before release. It is released under the Creative Commons Zero license.\\
\textbf{LibriSpeech}\cite{lbs} dataset is a large-scale corpus of around 1000 hours of English speech. The data has been sourced from audiobooks from the LibriVox project and is 60 GB in size. The data has been carefully segmented and aligned. It consists of 16kHz of reading English speech, prepared by Vassil Panayotov with the assistance of Daniel Povey. \\
\textbf{Ravdess}\cite{livingstone_steven_r_2018_1188976} is an emotional speech and song dataset. It made use of 24 professional actors to develop this dataset. This dataset was gender-balanced. Speech includes calm, happy, sad, angry, fearful, surprise, and disgust expressions, and the song contains calm, happy, sad, angry, and fearful emotions. They produced each expression with two levels of intensity, with an additional neutral emotion. They also validated the data by having around 247 unbiased individuals rate the dataset on its emotional standpoint. It was also released under a Creative Commons license. It is released under a Creative Commons BY 4.0 license.\\
\textbf{SAVEE}\cite{HaqEtAl_AVSP08}: Surrey Audio-Visual Expressed Emotion is an emotional database. It has recordings from 4 male actors in 7 different emotions with a total of 480 British English utterances. The sentences were picked from TIMIT corpus and phonetically-balanced for each emotion. It's an audiovisual dataset. The dataset is released under its own license "SAVEE DATABASE LICENCE AGREEMENT". The database is available free of charge for research purposes.\\
\textbf{Mivia Audio events}\cite{Foggia:2015:RDA:2829375.2829414} dataset is a dataset that consists of sound for surveillance applications. It has a total of 6000 events for surveillance applications, namely glass breaking, gunshots, and screams. This dataset emphasized on an amalgamation of noise caused by the distance between the actual noise source and the microphone. The data set is designed to provide each audio event at 6 different values of signal-to-noise ratio (namely 5dB, 10dB, 15dB, 20dB, 25dB, and 30dB) and superimposed on different combinations of environmental sounds to simulate their occurrence in different ambiances.

\section{Evaluation \& Results}
\label{sec:eval}
This dataset was created to provide the diversity in existing audio datasets allowing researchers to benchmark their models for keyword spotting and emotion classification. In this dataset, we also added a 3rd dimension, background sound, which give an extra edge to researchers identifying the background sounds.
Lastly, we have also provided the audio samples with labels of different place where the recordings were created.

For example, the audio's which were recorded in parks were stored separately while those taken in the house were stored in a different folder and the same was done for other places. This type of data compilation process helps in inverse mapping of background sounds to their environment of generation. The other purpose this solves is to make the model more robust as a large subset of the recording contains background sound which will further make the model work better in noisy environments.\\

The distribution of the dataset below is a reflection of the original dataset that we collected. We have also augmented our dataset with different background sounds such as that of public places and screams. This dataset will also be released along with original dataset.\\

\begin{figure}[ht]
    \centering
    \begin{tikzpicture}
\begin{axis}[
    x tick label style={
        /pgf/number format/1000 sep=},
    ylabel=No of Samples,
    enlargelimits=0.05,
    enlarge x limits=0.12,
    legend style={at={(0.5,-0.2)},
    anchor=north,legend columns=-1},
    ybar,
    ymin=0,
    symbolic x coords={Help, Bachao, Yes, No, Stop, Go}
]
\addplot 
    coordinates {(Help,1070) (Bachao,948)
         (Yes,877) (No,787) (Stop,788) (Go,965)};
\addplot 
    coordinates {(Help,3210) (Bachao,2844)
         (Yes,2631) (No,2361) (Stop,2364) (Go,2895)};
\legend{Before Augmentation, After Augmentation}
\end{axis}
\end{tikzpicture}
    \caption{This figure shows the change in no of training samples caused by data augmentation.}
    \label{fig:data_aug}
\end{figure}
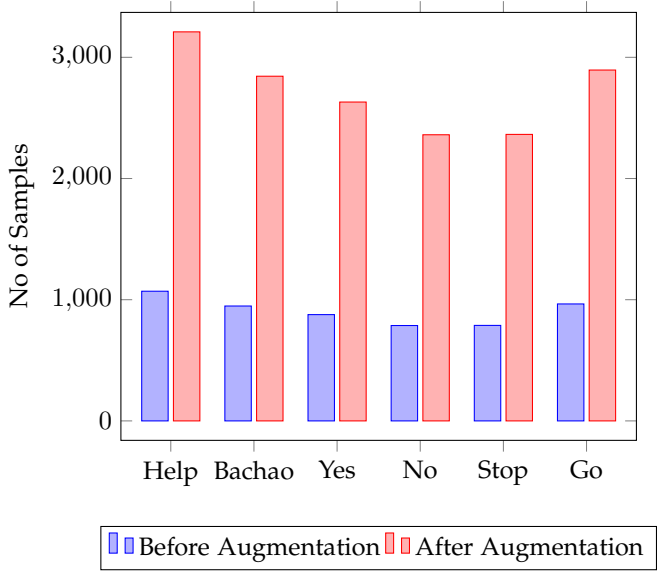

Tables \ref{table:keywords} and \ref{table:emotions} represent distribution of our dataset in keywords and emotion respectively. The fig. \ref{fig:data_aug} gives actual reflection of training data. This gives the number of samples of each class mentioned in table \ref{table:keywords}, table \ref{table:emotions} which are used while training. This is made by augmenting the collected data using speech augmentation techniques like white noise addition, shifting and stretching. These techniques are explained later in the section.
\begin{table}[ht!]
\centering
\begin{tabular}{||c c||} 
 \hline
 Word & Number of Utterances\\ [0.5ex] 
 \hline\hline
 Help & 1070 \\ 
 \hline
 Bachao & 948 \\
 \hline
 Yes & 877 \\
 \hline
 No & 787 \\
 \hline
 Stop & 788 \\
 \hline
 Go & 965 \\
 \hline
 Unknown & 1192 \\
%  \hline
%  Background & 1116 \\
 \hline
\end{tabular} 
\caption{Number of samples of each keyword}
\label{table:keywords}
\end{table}
% The table \ref{Figure:2} gives the number of samples of each emotion in our dataset.
\begin{table}[ht!]
\centering
\begin{tabular}{||c c||} 
 \hline
 Emotion & Number of Samples\\ [0.5ex] 
 \hline\hline
 Calm & 3826 \\ 
 \hline
 Fearful & 1630 \\
 \hline
 Happy & 988 \\
 \hline
 Angry & 1136 \\
 \hline
\end{tabular}
\caption{Number of samples of each emotion}
\label{table:emotions}
\end{table}

Then to get a balanced training set we augmented the keywords with background sound.\\
% \begin{center}\fbox{\textbf{Training Set contains 6000 samples of each keyword.}}\end{center}
In the absence of an adequate volume of training data, it is possible to increase the effective size of existing data through the process of data augmentation, which has contributed to significantly improving the performance of deep networks in the domain of image classification. In the case of speech recognition, augmentation traditionally involves deforming the audio waveform used for training in some fashion (e.g., by speeding it up or slowing it down), or adding background noise. This has the effect of making the dataset effectively larger, as multiple augmented versions of a single input is fed into the network throughout training, and also helps the network become robust by forcing it to learn relevant features. 

\subsection{Training Standards}
The standard chosen for training is as follows:
\begin{enumerate}
  \item The training data should be balanced in each class, that is, keywords, emotions and background samples should be balanced for training their respective models. 
  \item We used 3 types of speech augmentation namely additive white noise, stretch and wave shifting.
  \begin{itemize}
     \item \textbf{Noise Addition: }All of these affect the waveform differently. Adding white noise or augmenting some audio samples from background class to the actual audio. This operation is also referred to as 'noise injection' or 'noise corruption'. The noise-corrupted speech data are then used to train DNNs (Deep Neural Networks) as usual. The rationale of this approach is two-fold: firstly, the noise patterns within the introduced noise signals can be learned and thus compensated for in the inference phase, which is straightforward and shares the same idea as the multi-condition training approach; secondly, the perturbation introduced by the injected noise can improve generalization capability of the resulting DNN, which is supported by the noise injection theory.
     \item \textbf{Stretching: }It is the process of changing the speed or duration of an audio signal without affecting its pitch. It helps in the generalization of the variable compression and rarefaction phenomenon caused by sound vibration.
     \item \textbf{Shifting: }The idea of shifting time is very simple. It just shifts audio to left/right with a random second. If shifting audio to left (fast forward) with x seconds, first x seconds will mark as 0 (i.e. silence). If shifting audio to right (back forward) with x seconds, last x seconds will mark as 0 (i.e. silence).
   \end{itemize}
   \item We have made a train, validation and test split as follows: 70\%, 10\%, 20\% (Refer fig. \ref{fig:split}).
  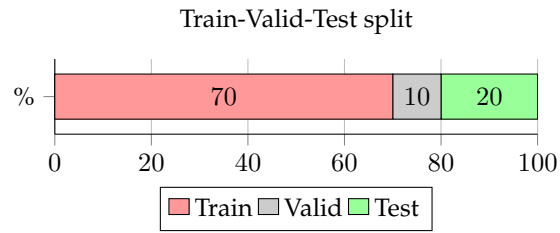
\begin{figure}[ht]
      \pgfplotsset{split/.style={
        title=Train-Valid-Test split,
        xbar stacked,
        width=8cm,
        axis y line*= none, axis x line*= bottom,
        xmajorgrids = true,
        xmin=0,xmax=100,
        ytick = data,
        yticklabels = {\%},
        tick align = outside, xtick pos = left,
        bar width=6mm, y=8mm,
        enlarge y limits={abs=0.625},
        nodes near coords,
        nodes near coords align={center},
        }, compat=1.14}
        \begin{tikzpicture}
        \begin{axis}[split, legend style={at={(0.5,-0.7)},
        anchor=north, legend columns=-1}] 
            \addplot[fill=red!40] coordinates{(70,0)};
            \addplot[fill=gray!40] coordinates{(10,0)};
            \addplot[fill=green!40] coordinates{(20,0)};
            \legend{Train, Valid, Test};
        \end{axis}
        \end{tikzpicture}
      \caption{Train-Valid-Test split}
      \label{fig:split}
  \end{figure}
\end{enumerate}

\subsection{Metrics}

\subsubsection{Top-One Error}
The Top-1 error is the percentage of the time that the classifier is not able to give the correct class the highest score. Now unlike image classification tasks where the model gives confidence score to one of the categories that it has been trained on, our classification model should be able to assign confidence score based on number of factors such as whether the audio contains silence or it contains the word that it does not recognize as one of the keyword.\\
This situation is dealt with, by introducing the background and unknown class. Each has its function like background class audio contains the sounds whose source is not human vocal chords or more precisely sounds which cannot be classified as words like birds chirping, roadside noise, or some random white noise as we experience in noisy places like crowded markets, underground subways etc, where we are not able to hear or identify any word in particular and if we are able to identify any word in our audio then it is classified as unknown.\\
This type of labeling captures the "open world" category in a very comprehensive manner. This makes the working of the model more robust as it need not always give an evenly distributed confidence score in response to unknown words or silence or any background sound.\\
We trained our dataset on a modified version of EdgeSpeechNet(B)\cite{Lin2018EdgeSpeechNetsHE} model which has proved its efficiency and effectiveness in training models with less number of category in the dataset which in our case is about 8 keyword class for keyword spotting and 4 emotion for emotion classification. This model also produced a very small model size that is around 2.2MB. This can easily be ported to mobile devices or any embedded devices and can be used in real-time.

This EdgeSpeechNet(B) model serves as baseline for this dataset and all the accuracy mentioned in Table \ref{table:accuracy} is based on top-one error metric. It gives the accuracy of EdgeSpeechNet(B) for different tasks that can be performed using our dataset.
\begin{table}[ht!]
\centering
\begin{tabular}{||c c||} 
 \hline
 Task & Accuracy\\ [0.5ex] 
 \hline\hline
 Keyword Spotting & 92\% \\ 
 \hline
 Emotion Classification & 90\% \\
 \hline
%  Background Classification & 88\% \\
%  \hline
\end{tabular}
\caption{Accuracy of EdgeSpeechNet on different tasks}
\label{table:accuracy}
\end{table}
\subsubsection{Streaming Error Metric}
The testing of samples as a segment of the audio clip also makes it different from the real-world scenario where the audio is perceived as a continuous stream. So in order to get a real understanding of working of the model, we have to test it on a stream of audio.\\
To generate an audio stream, we collected a 5-minute recording having different keywords in it and labeled it manually as a 3-second segment with 1-second stride.\\

We were able to achieve 52\% accuracy at classifying the speech samples.
.\\This result represents that in a recording of 5 minutes, 52\% of the samples, which are of 3 seconds each, are classified correctly. Unlike Speech Command, for this test metric, we didn't consider any tolerance between the time a label that is predicted by the model and when the actual label should have been predicted. This is because, in our case, there is ample overlap of the sliding window for consecutive test samples and hence tolerance time is not required. Speech command used the tolerance time concept because of relatively comparable sizes of sample duration and stride (both 1 second long).\\
\subsection{Comparison: Speech Command \& Indian EmoSpeech Command}

\begin{figure*}[ht]
    \centering
    \includegraphics[scale=0.55]{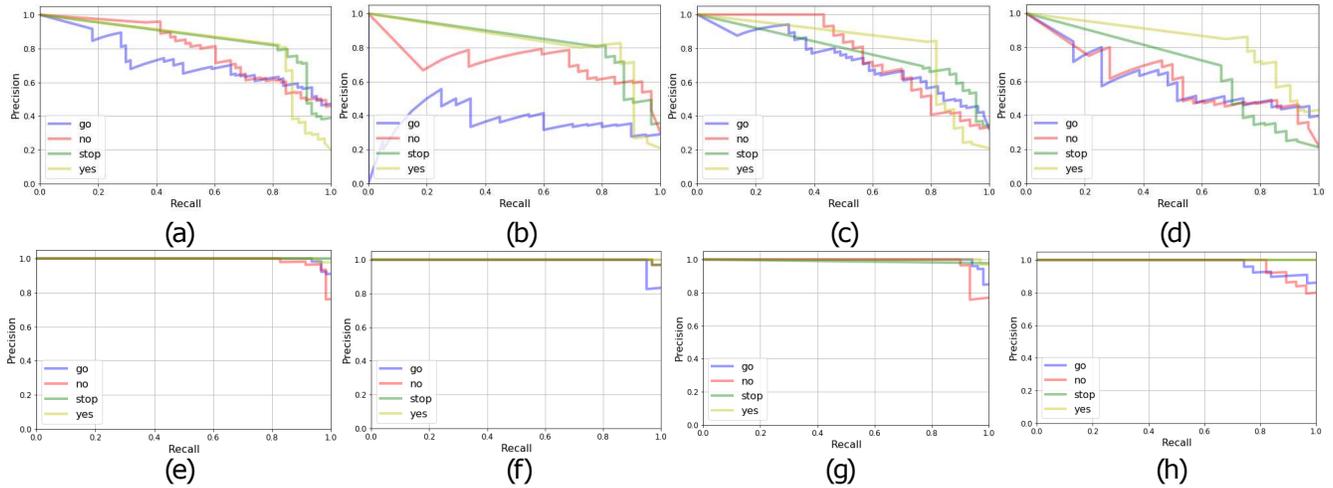}
    \caption{(a)-(d)Speech Command: PR Curve for keywords in i) Calm ii) Fearful iii) Angry iv) Happy Emotions. (e)-(h) Indian EmoSpeech Command: PR Curve for keywords in i) Calm ii) Fearful iii) Angry iv) Happy Emotions.}
    \label{fig:pr_speech_emo}
\end{figure*}
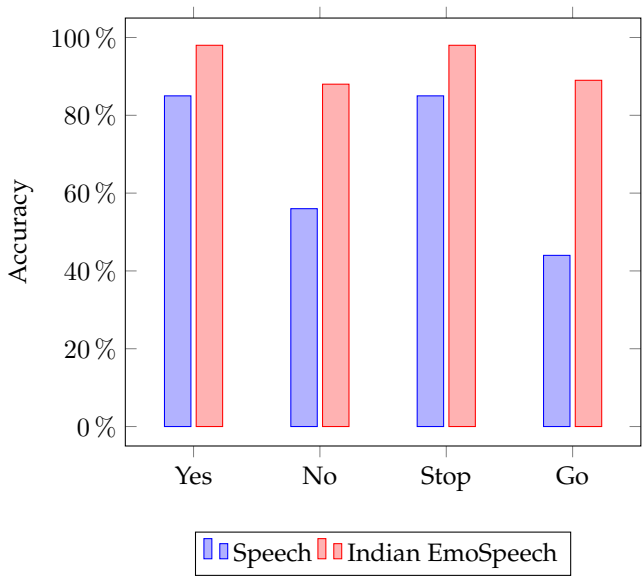
\begin{figure}[ht]
    \centering
    \begin{tikzpicture}
    \begin{axis}[
        x tick label style={
            /pgf/number format/1000 sep=},
        ylabel=Accuracy,
        enlargelimits=0.05,
        enlarge x limits=0.18,
        legend style={at={(0.5,-0.2)},
        anchor=north,legend columns=-1},
        ybar,
        ymin=0,
        ymax=100,
        yticklabel=\pgfmathprintnumber{\tick}\,$\%$,
        symbolic x coords={Yes, No, Stop, Go}
    ]
    \addplot 
        coordinates {(Yes,85) (No,56) (Stop,85) (Go,44)};
    \addplot 
        coordinates {(Yes,98) (No,88) (Stop,98) (Go,89)};
    \legend{Speech, Indian EmoSpeech}
    \end{axis}
    \end{tikzpicture}
    \caption{The figure shows a comparison between the model performance of Speech Command model and Indian EmoSpeech Command model.}
    \label{fig:acc_all}
\end{figure}

This section covers the comparison in the performance of the models trained on Speech command (baseline model) and the Indian EmoSpeech command. It is indicative of the gaps that existing speech command dataset currently have and how the Indian EmoSpeech Command meets those gaps while providing several application use cases.\\
The disparity between the Indian EmoSpeech command and Speech command datasets is the presence of emotion modality. This modality expands its application domain to a large extent. Secondly, it also provides a more diverse dataset to work on. The emotion part of the Indian EmoSpeech Command dataset enables it to provide a keyword in diverse speech patterns while facilitating emotion analysis along with speech. This makes keywords in the Indian EmoSpeech Command dataset very different from the neutral utterances of keywords of the Speech Command dataset. All of this is evident in our analysis of Speech Command and Indian EmoSpeech Command dataset. The process of evaluation is as follows:
\begin{enumerate}
    \item A model is trained on some Speech Command keywords like "go", "no", "stop", "yes" which are also present in the Indian EmoSpeech Command dataset. The speech Command dataset has around 2300 keywords for each of these keywords.
    \item This model is then tested on the Indian EmoSpeech Command test set containing the above-mentioned keywords.
    \item The accuracy numbers for the trained model as mentioned above is compared against the model trained on the same keywords of the Indian EmoSpeech Command dataset. 
    \item We use the precision and recall as the metrics for evaluation, which are calculated as follows:\\
     \[Precision = \frac{TP}{TP+FP}\]
     \[Recall = \frac{TP}{TP+FN}\]
  where TP represents True Positives and it means when a sample is classified correctly as Positive, FP represents False Positives and it means when a sample is classified incorrectly as Positive, FN represents False Negatives and it means when a sample is incorrectly classified as Negative.
  \\Precision-Recall (PR) curves are then plotted to perform a comprehensive analysis and to understand the confidence of the model trained on Speech Command keywords. This analysis is based on different emotions, namely, "calm", "fearful", "angry" and "happy". As we can see in table \ref{table:keywords} the number of samples for each keyword is not very high. Also, we are only using 20\% of them for building the test set. So, when working with one vs all classification inference, we will have an abundance of negative samples. Therefore, the PR curve is more suitable for such inference.
\end{enumerate}
It is important to note that the model architecture used for training on both the dataset is the same, that is a modified version of EdgeSpeechNet(B). Also, all the hyperparameters are same as used in Speech Command Model. We utilize adam Optimizer for learning and minimizing the loss function. The loss function used for the classification models is categorical cross-entropy given as,
\[H_{y'} (y) := - \sum_{i} y_{i}' \log (y_i)\]
\\where y' is predicted confidence score on overall classes and y is actual confidence score on overall class.
\\\\We compared the 2 trained models, which are models trained on Speech command and Indian EmoSpeech command, were compared based on top-one accuracy metric. The figure \ref{fig:acc_all} shows the comparison between these two models.
\\These accuracy numbers present a piece of very crucial evidence. Despite having fewer true samples for each of these keywords to train on for the Indian EmoSpeech Command model, still, it performs better than the Speech Command model. It validates the diversity the Indian EmoSpeech Command dataset has in terms of background sound and emotion. This diversity helps in a better generalization of the keywords. This is what the speech command dataset lacks and hence performed miserably compared to the Indian EmoSpeech Command model.
\\The emotion-wise analysis presents a much more comprehensive analysis of the Speech Command model, the model trained on the Speech Command dataset, which will help us inference the working of the model on different speech patterns that are influenced by emotions.

\subsubsection{Analysis on Calm Emotion}
The fig. \ref{fig:pr_speech_emo}(a) (Speech Command: Calm) and \ref{fig:pr_speech_emo}(e) (Indian EmoSpeech Command: Calm) shows how a different accent and change in voice caused by emotion association with keyword makes the model trained on it more robust. As can be seen in fig. \ref{fig:pr_speech_emo}(a) these diverse test samples caused a deterioration of performance in the Speech Command model. For high decision thresholds, the confidence score for the prediction of the keyword decreases steadily. But for the same test data, the Indian EmoSpeech Command model performed well enough to identify the keywords with a high confidence score.

 \subsubsection{Analysis on Fearful Emotion}
Fig. \ref{fig:pr_speech_emo}(b) (Speech Command: Fearful) and \ref{fig:pr_speech_emo}(f) (Indian EmoSpeech Command: Fearful) interpretation is as follows. For Indian EmoSpeech Command model at a higher decision threshold, the keyword "go" has a low confidence score than others in case of fearful emotion. 
The speech command model has a low confidence score even at a low decision threshold. This suggests classifier is very selective towards other keywords in the samples with fearful emotion and rarely classifies keywords as "go". The same is the case for "no" but less appreciable than "go". "stop" and "yes" as usual performs well at a low decision threshold and then sharply falls at a high decision threshold.
\subsubsection{Analysis on Angry Emotion}
Fig. \ref{fig:pr_speech_emo}(c) (Speech Command: Angry) shows that the Speech Command model to have steadily decreasing precision as seen in calm emotion. We can also see "stop", which had shown good performance for other emotions, has fewer positive predictions that are true at higher recall. This is due to varying speech patterns within angry emotion bracket. In comparison, the Indian EmoSpeech Command model (fig. \ref{fig:pr_speech_emo}(g) (Indian EmoSpeech Command: Angry)) tries to learn the general features of the angry emotion of voice and was successful in identifying the said keyword. Hence this comparison suggests stress and emotion affect your voice and make words sound differently and so having such data makes keyword recognition better in various scenarios.

\subsubsection{Analysis on Happy Emotion}
The Fig. \ref{fig:pr_speech_emo}(d) (Speech Command: Happy) and Fig. \ref{fig:pr_speech_emo}(h) (Indian EmoSpeech Command: Angry) again revalidates the inferences made above. The different emotions make words sound different and make their spectrum have different amplitudes for the same frequencies. In our "happy" emotion, keywords are pronounced with laughs and giggles. And hence we see a fall in the precision score of all the keywords at higher recall. Here, we can see the PR curve of "go" and "no" began to fall even before others this was due to the phonetic similarity in them and so these become harder to distinguish between as giggles and laughs were added to it.

\section{Application}
\label{sec:appli}
Given the construction of this dataset, it can find usage in audio surveillance tasks. The keywords that the dataset contains completely tackle the problem of security. The addition of emotions and background sounds further extend its serviceability by allowing to analyze the speaker's emotional state from the speech itself and helping in identifying the environment, surrounding possible location. A recognition model trained on noises can also help identify the condition of the environment in the vicinity of the receiver.
\\ Since this dataset also contains the background noise in a large subset of its samples, it can help in development of models that segregates foreground sound and background sound effectively.
\section{Future Work}
\label{sec:future}
Considering the applications and the diversity we have two goals for the future of the Indian EmoSpeech command dataset:
\subsection{Adding more keywords}
We are looking forward to adding more keywords to this dataset. This will increase the applications of the dataset. We will be using the same platform as we have mentioned earlier.
\subsection{Adding more emotions}
We will also add more emotions to the dataset in the upcoming time. This will make it a comprehensive dataset for emotion recognition and analysis in speech. Psychological studies could benefit from using this dataset to create models to interpret emotional state from the person's speech. This is based on the fact that there is a clear relation between emotional state of person and their speech. This is explained through physiological response triggered throughout body due to psychological state of mind. For instance, when a person is angry, their breathing becomes faster, blood pressure rises, jaw becomes clenched, mouth becomes dry, pupils are dilated and their overall circulatory system is under stress. Due to all these changes as compared to normal conditions, person's angry voice is distinguishable from their neutral voice. Also, research shows our voice often reveals even more than our faces in terms of our intended message.%[https://greatergood.berkeley.edu/article/item/does_your_voice_reveal_more_emotion_than_your_face].
Therefore, adding more emotions make it suitable for analysis of the state of mind from the voice itself.
\subsection{Exploiting the dataset for adversarial attacks}
The background sound in the Indian EmoSpeech Command dataset may cause some examples to act as adversarial examples to our speech recognition model. And hence we will also be looking for possible defenses and methods that could help us identify those examples\cite{gideon2019improving,abdelwahab2018domain}.

\section{Conclusion}
\label{sec:conclu}
We have presented an indigenous Indian EmoSpeech Command dataset which has been shown to have potential in safety and emotion detection applications. It is a 3-dimensional dataset in terms of the information contained in it in terms of keywords, emotions and background noise. We have demonstrated that the dataset performs significantly well in emotion recognition as compared to already existing speech command datasets. We have shown significant average gain of 3.3\% in top-one score over a subset of speech command dataset for keyword spotting.  

\ifCLASSOPTIONcompsoc
  \section*{Acknowledgments}
\else
  \section*{Acknowledgment}
\fi

Many thanks are owed to everyone who contributed with their voice samples for this dataset. We are extremely grateful for each contribution. We couldn't have put together this research project without the help and support of Dr. Aakanksha Chowdhery of Google Brain, Prof. Brejesh Lal of IIT Delhi and the team of Celestini Project India. We are most grateful to these mentor figures.

\ifCLASSOPTIONcaptionsoff
  \newpage
\fi



\bibliographystyle{IEEEtran}
\bibliography{main}
\end{document}